\documentclass[12pt]{iopart}

\usepackage{graphicx}
\usepackage{color}

\begin{document}

\title[Complex Dirac Delta]{The complex Dirac Delta, Plemelj formula, and integral representations}

\author{J. Julve}

\address{IMAFF, Consejo Superior de Investigaciones
Cient\'\i ficas, Serrano 113 bis, Madrid 28006, Spain}

\ead{julve@imaff.cfmac.csic.es}

\author{R. Cepedello}

\address{Universitat de Valencia, Burjassot (Valencia), Spain}

\ead{ricepe@alumni.uv.es}

\author{F. J. de Urr\'\i es}

\address{Universidad de Alcal\'a de Henares, Spain}

\ead{fernando.urries@uah.es}

\begin{abstract}
The extension of the Dirac Delta distribution (DD) to the complex field
is needed for dealing with the complex-energy solutions of the
$\rm Schr\ddot{o}dinger$ equation, typically when calculating their
inner products.
In quantum scattering theory the DD usually arises as an integral
representation involving plane waves of real momenta.
We deal with the complex extension of these representations by using
a Gaussian regularization. Their interpretation as distributions requires
prescribing the integration path and a corresponding space of test functions.
An extension of the Sokhotski-Plemelj formula is obtained. This definition
of distributions is alternative to the historic one referred to surface
integrations on the complex plane.


\end{abstract}

\maketitle

\section{Introduction}

We consider 1-dimensional quantum barriers of compact support,
often referred to as cut-off potentials,
and the solutions to the corresponding time-independent
$\rm Schr\ddot{o}dinger$ equation.
Besides the real-energy solutions, namely the bound states
(negative discrete spectrum) and the scattering states (positive
continuum spectrum) or "Dirac kets", of well-known physical meaning,
there exists a much larger set of complex-energy solutions. These
solutions may be obtained by analytic continuation
of the Dirac kets to the whole complex plane. When so doing, a discrete set of resonant solutions
(Gamow states) arises embedded in the sea of the general "background solutions" \cite{Nussenzveig}.
Asymptotically the Gamow states feature only either outgoing or incoming
plane waves with (complex) momenta corresponding to poles of the S-matrix.
We notice that in the complex-momentum plane (which is isomorphic to a
two-Riemann-sheet complex-energy plane), the bound states correspond to
momenta in the positive imaginary axis, their mirror momenta in the negative
imaginary axis corresponding to the anti-bound states. Both bound and anti-bound
momenta are also poles of the S-matrix. One should also notice that the resonances,
the background and the anti-bound solutions exhibit an exponential spatial growth
that places them far outside the Hilbert space ${\cal L}^2$ .

In the calculation of inner products between states of the continuum spectrum with real momentum,
integrals of the type

\begin{equation}
I(\pm k)\equiv\int^{+\infty}_0 \rmd x\; \rme^{\pm\rmi\,kx }\hskip 1.0cm (k\in {\cal R})
\end{equation}

\noindent{arise} \cite{JulveUrries}\cite{JulveTurriniUrries}. These integrals can be regularized by adding to $k$ a small imaginary part
$\pm\rmi\epsilon$, which ensures the convergence of (1), and then letting $\epsilon\rightarrow 0^+$.
The result is meaningful as a distribution \cite{Gelfand}
given by the well-known Sokhotski-Plemelj formula

\begin{equation}
I(\pm k)=\frac{\pm\rmi}{k\pm\rmi\,0^+} =\pm\rmi\,PV\frac{1}{k}+\pi
\,\delta(k)\hskip 0.2cm.
\end{equation}
For freely propagating plane waves with momenta $p$ and $p'$, the simpler integral

\begin{equation}
\int^{+\infty}_{-\infty} \rmd x\; \rme^{\rmi kx }=I(-k)+I(k)\hskip 1.0cm (k=p-p')
\end{equation}
arises and yields the usual result $I(-k)+I(k)=2\pi\delta(k)$.

Continuing $I(k)$ in (1) to the complex plane, namely defining $I(z)\;\;(z\in{\cal C})$, requires
further regularization of the integral as long as it generally diverges for ${\rm Im}\,z<0$.
We adopt the widely used Gaussian regularization \cite{Zel'dovich}\cite{Berggren}

\begin{equation}
J(z,\lambda)\equiv\int^{+\infty}_0 \rmd x\; \rme^{-\lambda
x^2}\rme^{\rmi zx }=\frac{\rmi}{z}\;\sqrt{\pi}\,w\,\rme^{w^2} {\rm
erfc}(w)\hskip 1.0cm (\lambda\;{\rm real}>0)
\end{equation}
which is directly related to (7.1.2) in \cite{Stegun}, where
$w=-\rmi z/(2\sqrt{\lambda})$. The limit $\lambda\rightarrow 0^+$ is taken afterwards.
Notice that, when completing a square in the exponent of the integrand in (4) and
changing the integration variable to $t=\sqrt\lambda\;x+w$ , one
obtains the intermediate expression
\begin{equation}
J(z,\lambda)= \frac{1}{\sqrt\lambda}\;\rme^{w^2}\int^{\infty+w}_w
\rmd t\; \rme^{-t^2}
\end{equation}
The integral representation (7.1.2) is convergent when the path of
the complex integration variable $t$ approaches $\infty\;$ along a
direction $-\frac{\pi}{4}<{\rm arg}(t)<\frac{\pi}{4}$ . This
condition is fulfilled by (5) for any finite $w$.
However, in (4) the limit $\lambda\rightarrow 0^+$ (and
hence $w\rightarrow\infty$) not always can be taken before the
integration is carried out, because the function
$f_{\lambda}(x)\equiv\rme^{-\lambda x^2}$ does not converge
uniformly to the function $f_0(x)=1$ when $\lambda\rightarrow 0^+$ .

For Im $z\;>0$ the limit can be taken in the integrand in (4)
because the finiteness of the integral is always assured by the
convergence factor $\rme^{-({\rm Im}\,z)\,x}$, trivially yielding
$\rmi\,z^{-1}$. In this case $J(z,\lambda)$ converges to
$J(k,0)=\rmi z^{-1}$ when $\lambda\rightarrow 0^+$ .

For real $z$ the result (2) is obtained.

For Im $z\;<0$ the integration and the limit $\lambda\rightarrow 0^+$
do not commute, in which case we adopt the limit of the integral as a
prescription. In the right-hand side of (4), the limit
$\lambda\rightarrow 0^+$ can be directly inferred from (7.1.23) in
\cite{Stegun}, namely
\begin{equation}
{\rm lim}_{w\rightarrow\infty}\sqrt{\pi}\,w\,e^{w^2}\,{\rm erfc}(w)=
\left\{
\begin{array}{ll}
1&\quad ,\quad -3\frac{\pi}{4}< {\rm arg}(w)<3\frac{\pi}{4}\\
\infty & \quad ,\quad {\rm otherwise}
\end{array} \right.
\end{equation}
Therefore (4) yields an extension of $J(z,0)$ to a new region of the
lower half complex plane, namely
\begin{equation}
 J(z,0)=\left\{
\begin{array}{ll}
\frac{\rmi}{z}&\quad ,\quad
-\frac{\pi}{4}<{\rm arg}(z)<5\frac{\pi}{4} \;\;\;,\;\; k\neq 0\\
\infty & \quad , \quad{\rm otherwise}
\end{array}
\right.
\end{equation}
The domain ${\cal D}_+$ where $J(z,0)$ is finite is thus the complex plane $\cal{C}$ minus
a wedge in the lower half plane that includes the point $z=0$. We call $J(z,0)\equiv I(z)$ for short.

\section{Distribution in the complex plane}

Interpreting $I(k)$ in (2) as a distribution means that

\begin{equation}
\int^{+\infty}_{-\infty} \rmd k\; I(k)\,f(k)= \rmi\,PV\int^{+\infty}_{-\infty} \rmd k\; \frac{1}{k}\,f(k)\; +\; \pi\,f(0)
\end{equation}
on a set of test functions $f(k)$. Notice that the integration on the real axis $k\in{\cal R}$ is involved in this definition.

Likewise, interpreting (7) as a distribution requires prescribing a set of test functions and an integration path in the complex plane. Consider a (finite or infinite) path $\Gamma \subset {\cal D}_+$ and the set of analytic functions $f_\Gamma(z)$ which are integrable along $\Gamma$. Then

\begin{equation}
\oint_{\Gamma} \rmd z\; I(z)\,f_\Gamma(z)= \rmi\oint_{\Gamma} \rmd z\; \frac{f_\Gamma(z)}{z} \quad<\infty \quad ,
\end{equation}
and (9) defines a functional trivially involving $I(z)= \rmi z^{-1}$ in a strict function sense.

We now show that, for paths $\Gamma\subset{\cal D}_+$ albeit for crossing the singular point $z=0$,  $I(z)$ still has a meaning as a distribution in the enlarged sense above\footnote[1]{In the real axis, precise conditions on the functions $f(z)$ for (9) to exist can be found in \cite{CZhu}.}. In fact, we can divert $\Gamma$, so as to fully keep it within ${\cal D}_+$, by deforming it to a new path $\Gamma_\epsilon + {\cal C}_\epsilon$, where $\Gamma_\epsilon$ is the path $\Gamma$ stripped of a segment of length $2\epsilon$ centered in $z=0$, ${\cal C}_\epsilon$ is a half-circle of radius $\epsilon$ connecting the two parts of $\Gamma_\epsilon$ and circling $z=0$ from above, and then letting $\epsilon\rightarrow 0^+$ (Fig.1). This is equivalent to shifting the path $\Gamma$ upwards by the amount $+\rmi\epsilon$.

\begin{figure}[h]
\begin{center}
\includegraphics[width=0.8\textwidth]{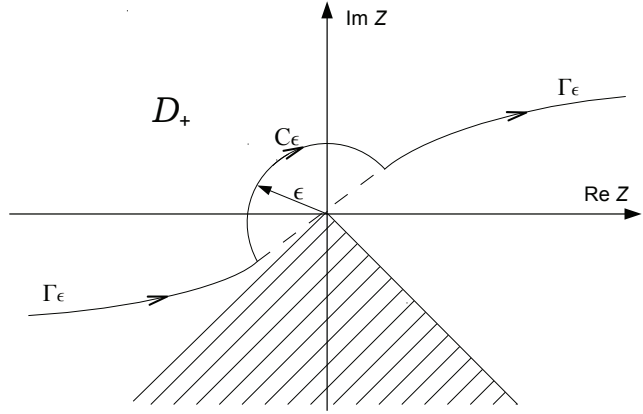}
\caption{Deformation of the integration path $\Gamma$ \hfill\break where the set of test functions is defined.}
\label{Fig.1}
\end{center}
\end{figure}

\noindent{The} result
\begin{equation*}
\hspace*{-1.5cm}
{\rm lim}_{\epsilon\rightarrow 0^+}\{\oint_{\Gamma_\epsilon} \rmd z\; I(z)\,f_\Gamma(z)+\oint_{{\cal C}_\epsilon} \rmd z\; I(z)\,f_\Gamma(z)\}= \rmi\;PV\oint_{\Gamma} \rmd z\; \frac{1}{z} f_\Gamma(z)\;+\;\pi f_\Gamma(0)
\end{equation*}
\noindent{lets} us interpreting
\begin{equation}
I(z)=\frac{\rmi}{z+\rmi\,0^+}=\rmi\,PV\frac{1}{z}+\pi
\,\delta(z)
\end{equation}
as a distribution for integration paths $\Gamma \subset {\cal D}_+$ and crossing $z=0$. Notice also that,
according to (7), for $\epsilon>0$ the apex of the divergence wedge of $I(z)$ is at the point $-\rmi\epsilon$ .

Similarly, for paths $\Gamma$ in the domain ${\cal D_-}$ ($\equiv {\cal C}$ minus a wedge in the upper half plane containing $z=0$) and crossing also $z=0$ we have
\begin{equation}
I(-z)=\frac{-\rmi}{z-\rmi\,0^+}=-\rmi\,PV\frac{1}{z}+\pi
\,\delta(z)\quad.
\end{equation}

The complex generalization of (3) is equally clear-cut. For $\epsilon>0$ the apex of the divergent wedge of $I(-z)$ is at the point $+\rmi\epsilon$, so that the regular domain of $I(-z)+I(z)$ contains the point $z=0$, although in the limit $\epsilon\rightarrow0^+$ the wedges pinch this point and the domain ${\cal D}={\cal D}_-\cap{\cal D}_+$ again excludes $z=0$ (Fig.2).

\begin{figure}[h]
\begin{center}
\includegraphics[width=0.8\textwidth]{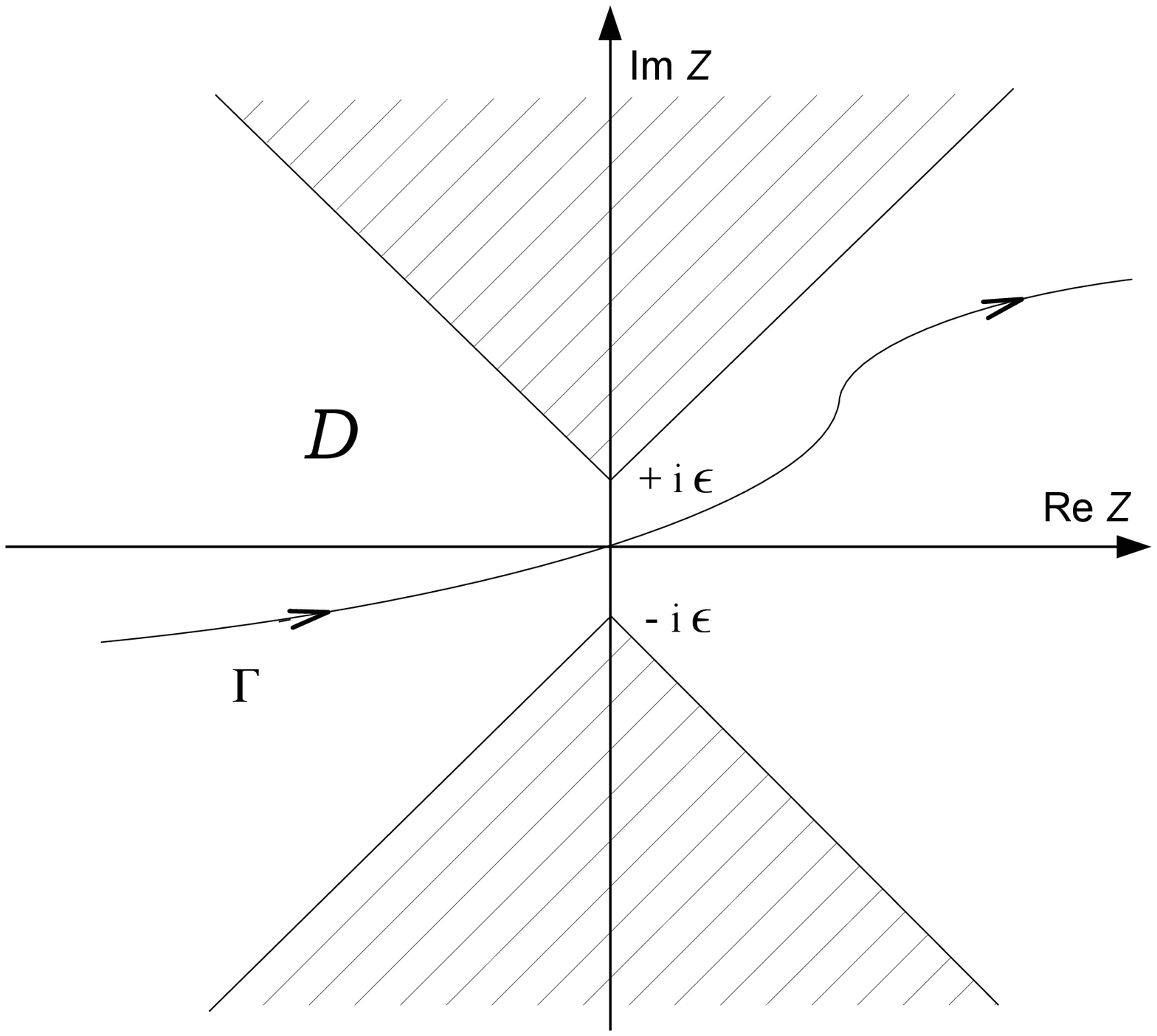}
\caption{Integration path crossing the point $z=0$ \hfill\break and avoiding the divergence wedges.}
\label{Fig.1}
\end{center}
\end{figure}

\noindent{We} then have
\begin{equation}
\int_{-\infty}^{+\infty}\rmd x\; \rme^{\rmi zx }=I(-z)+I(z)=2\pi\,\delta(z)
\end{equation}
for paths $\Gamma \subset {\cal D}$ and crossing $z=0$ when going from the left half
to the right half of the complex plane (which are a reminder of the real $-\infty$ to $+\infty$ integration).

\section{A derivation with real integrations}

The results in the previous sections can be attained along the more traditional lines of derivation
of the Sokhotski-Plemelj formula involving real integrations \cite{Gelfand}. This provides a complementary insight
that may be useful in some applications.

We re-derive the formula, as displayed in (2), albeit for the real line ${\cal R}$ tilted by a fixed angle $\phi$
in the complex plane, namely by writing $k=q\,\rme^{\rmi\phi}\,\in\,{\cal C}$, where $q\in{\cal R}$ and $-\pi/4<\phi<\pi/4$ initially (Fig.3).

\begin{figure}[h]
\begin{center}
\includegraphics[width=0.8\textwidth]{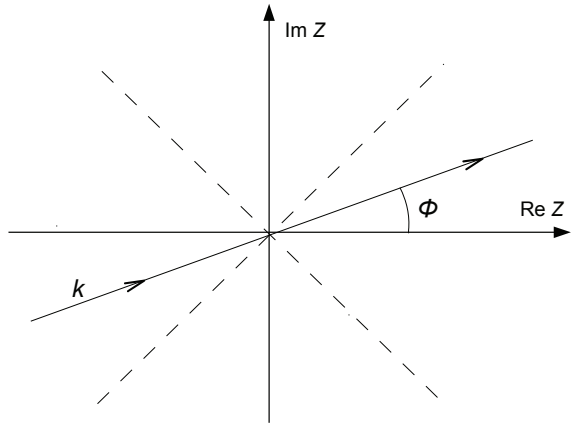}
\caption{The real integration path tilted in the complex plane.}
\label{Fig.1}
\end{center}
\end{figure}

\noindent{We} again add an imaginary part $\rmi\epsilon$ to $k$, so that
\begin{equation}
I(k+\rmi\epsilon)=\frac{\rmi}{k+\rmi\epsilon}
\end{equation}
and interpret it as a distribution when letting $\epsilon\rightarrow 0^+$. Again we write the denominator of (13) in polar form
\begin{equation}
k+\rmi\epsilon=|k+\rmi\epsilon|\;\rme^{\rmi\,{\rm arg}(k+\rmi\epsilon)}
\end{equation}
where, for the complex $k$ defined above,
\begin{equation}
{\rm arg}(k+\rmi\epsilon)={\rm arctan}\,\frac{q\,{\rm sin}\,\phi+\epsilon}{q\,{\rm cos}\,\phi}+\pi\,\Theta(-q)\quad.
\end{equation}
The first term in the right-hand side of (15) has a discontinuity at $q=0$ that just compensates for the one of the step function $\Theta(-q)$, so that (15) is a continuous function of $q$ , ranging from $\phi+\pi$ to $\phi$ for $q$ going from $-\infty$ to $+\infty$ , and becoming discontinuous for $\epsilon\rightarrow0^+$  , namely
\begin{equation}
{\rm arg}(k+\rmi\,0^+)=\phi+\pi\,\Theta(-q)\quad.
\end{equation}
Taking the logarithm of (14) we have
\begin{equation}
{\rm ln}(k+\rmi\epsilon)={\rm ln}|k+\rmi\epsilon|+\rmi\,{\rm arg}(k+\rmi\epsilon)
\end{equation}
and differentiating it with respect to $k$ we obtain
\begin{equation}
\frac{1}{(k+\rmi\epsilon)}=\frac{\rm d}{{\rm d}k}\,{\rm ln}|k+\rmi\epsilon|+\rmi\,\frac{\rm d}{{\rm d}k}\,{\rm arg}(k+\rmi\epsilon)\quad .
\end{equation}
The derivatives in (18) can be worked out by taking into account that $\frac{\rm d}{{\rm d}k}=\rme^{-\rmi\phi}\frac{\rm d}{{\rm d}q}$, but they are cumbersome and of little interest.
Taking the limit $\epsilon\rightarrow0^+$ and recalling (16) we obtain the simple expression
\begin{equation}
\frac{1}{k+\rmi 0^+}=\;\rme^{-\rmi\phi}PV\frac{1}{q}-\rmi\pi\,\rme^{-\rmi\phi}\frac{\rm d}{{\rm d}q}\Theta(-q)=PV\frac{1}{k}-\rmi\pi\,\delta(k)\quad .
\end{equation}
Albeit for the constant factor $\rme^{-\rmi\phi}$, (19) is the Sokhotski-Plemelj formula for the
real variable $q$, being therefore meaningful as a distribution on test functions $f(q\,\rme^{\rmi\phi})$.

The derivation of (19) above is actually valid for $-\pi/2<\phi<\pi/2$ , namely for straight
integration paths crossing the origin and sweeping almost the whole complex plane.
However (13) holds only for $k+\rmi\epsilon\in{\cal D}_+$ , that is, for $-\pi/4<\phi<\pi/4$ .

\section{Conclusions}

Extending the usual Dirac delta orthogonality between eigenstates of the continuum real energy and momentum to the so called "background solutions", which are the continuation of the former to the complex momentum plane, entails a corresponding complex generalization of the Dirac Delta (DD) distribution. This can be done by adopting a Gaussian regularization of the integral representation, arising when computing inner products generally involving background and resonant solutions. In fact, the Gaussian convergence factor is able to cope with the space exponential growth of these solutions.

We have seen that the distributions Cauchy Principal Value (CPV) of $\frac{1}{x}$ and the DD, originally defined for real variables, admit an extension to the complex plane, namely $PV \frac{1}{z}$ and $\delta(z)$, on the space of test functions $f_\Gamma(z)$ which are analytic and integrable on paths $\Gamma$ crossing the point $z=0$. Under these definitions, the Sokhotski-Plemelj (SP) formula has a natural extension to the whole complex plane. However, the Gaussian regularization of the usual integral representations yields the SP formula only for integration paths $\Gamma\subset{\cal D}={\cal D}_+\cap{\cal D}_-\subset{\cal C}$, where the domains ${\cal D}_+$ and ${\cal D}_-$ are the complex plane ${\cal C}$ minus wedge-shaped regions where the integrals diverge upon removing the regularization. This is the main result of this paper. Regularizations more powerful than the Gaussian one might exist that yield a regular domain even larger than ${\cal D}$, perhaps almost the whole ${\cal C}$.

The above definition of the extension of generalized functions to the complex plane, referred to line integrals of test functions, is alternative to the traditional one by Gel'fand et al. (see the Vol.I, App. B of \cite{Gelfand}), in which the generalized functions act on the test functions through surface (or $2m$-volume, for $m$ variables) complex integrals with the differential form $\rmd z\,\rmd \bar{z} = -2\rmi\,\rmd x\,\rmd y$. In that case the SP formula might not have any immediate and useful meaning. Thinking the complex extension of distributions in terms of line integrals rather than of surface integrals was already implicit in \cite{CZhu}, and is more suited for some problems of scattering theory.

In fact, an application of the above result is the following: The scattering solutions (continuum spectrum of real positive energy $E$) for a potential barrier are mutually DD-orthogonal and, if no bound states occur, also yield a complete basis of the Hilbert space. In any case, the operator $\int{\rm d}E\,|E\rangle\langle E|$ is idempotent (i.e. it is a projector) accordingly.
Consider now paths $\Upsilon\subset{\cal C}$ with maximum slope within the interval $(-\pi/4,+\pi/4)$, and the set of background solutions with momenta $z\in\Upsilon$, so that $z-z'\subset{\cal D}$ for any $z'\in\Upsilon$ else. Then our result leads to an analogous DD-orthogonality $\langle z|z'\rangle=\delta(z-z')$ between these background states\footnote[2]{The proper continuation of the bras, or "left kets", to the complex plane must be used \cite{Madrid2}.}, and hence $\oint_{\Upsilon}{\rm d}z\,|z\rangle\langle z|$ is a projector as well. This turns useful in the search for expansions of wave functions and operators in terms of resonances and background states \cite{Berggren2} \cite{Muga}.

\ack{J. Julve was supported by MINECO project FIS2011-29287}


\section*{References}
\begin{thebibliography}{13}

\bibitem{Nussenzveig} Nussenzveig H M 1972 {\it Causality and dispersion
relations} (New York and London: Academic Press)

\bibitem{JulveUrries} Julve J and Urries F J 2010 {\it J. Phys. A: Math. Theor.} {\bf 43} 175301

\bibitem{JulveTurriniUrries} Julve J, Turrini S and Urries F J 2010 {\it Int. J. Theor. Phys.} {\bf 53} 971

\bibitem{Gelfand} Gel'fand I M and Shilov G E 1964 {\it Generalized functions} (New York and London: Academic Press)

\bibitem{Zel'dovich} Zel'dovich Ya B 1961
{\it Sov. Phys. JETP} {\bf 12} 542

\bibitem{Berggren} Berggren T 1968
{\it Nucl. Phys.} A {\bf 109} 265

\bibitem{Stegun} Abramowitz M and Stegun I A 1965 {\it Handbook of Mathematical
Functions} (New York: Dover)

\bibitem{CZhu} Chang Zhong Zhu {\it A Sokhotski-Plemelj Type Formula} (unpublished)

\bibitem{Madrid2} de la Madrid R 2009
{\it SIGMA} {\bf 70} 626

\bibitem{Berggren2} Berggren T 1982
{\it Nucl. Phys.} A {\bf 389} 261

\bibitem{Muga} de la Madrid R, Garc\'{\i}a-Calder\'on G and Muga G
2005 {\it Czechoslovak Journal of Physics} {\bf 55} 1141







\end{thebibliography}
\end{document}